\documentclass[preprint]{aastex}
\usepackage{graphicx}
\usepackage[latin1]{inputenc}

\shorttitle{Position displacement of diffuse interstellar bands}

\shortauthors{Galazutdinov et al.}

\begin{document}

\title{Position displacement of diffuse interstellar bands
\thanks{This paper includes data gathered with the 6.5 meter Magellan Telescopes
located at Las Campanas Observatory(Chile).
}}

\author{G.~Galazutdinov$^{1,2}$}
\affil{$^1$ Instituto de Astronomia, Universidad Catolica del Norte, Av. Angamos 0610, Antofagasta, Chile}
\affil{$^2$ Pulkovo Observatory, Pulkovskoe Shosse 65, Saint-Petersburg 196140, Russia}
\email{runizag@gmail.com}

\author{J.~Kre{\l}owski }
\affil{Center for Astronomy, Nicholas Copernicus University,
Gagarina 11, Pl-87-100 Toru{\'n}, Poland}
\email{jacek@astri.uni.torun.pl}

\author{Y.Beletsky}
\affil{Las Campanas Observatory, Carnegie Observatories, Casilla
601, La Serena, Chile}
\email{ybelets@gmail.com}

\and

\author{G.Valyavin}
\affil{Special Astrophysical Observatory, Nizhnij Arkhyz, Russia}
\email{gvalyavin@gmail.com}

\begin{abstract}
We re-consider the already published phenomenon: the blue shift of
diffuse interstellar bands, observed in spectra of HD34078
(AE Aur) and members of the Sco OB1 association, in particular HD152233. We have
analyzed 29 diffuse bands. Part of them, already proven as
blue-shifted in our earlier study, are now confirmed using another
instrument: this time the 6.5m Clay telescope
equipped with the MIKE spectrograph.  The high
signal-to-noise ratio (over 600) of our spectra allowed us to
reveal even small small scale {\bf displacements} of positions (both: blue and
red-shifts) of diffuse bands along the considered lines of sight.
In some cases the magnitude of deviation exceeds 10 km\,s$^{-1}$.
Also, we prove that profiles of many diffuse bands in spectra of HD34078 suffer
significant broadening. The origin of the observed phenomena is discussed.
\end{abstract}

\keywords{ISM: lines and bands}

\section{Introduction}

Diffuse interstellar bands (DIBs) have been known since the
discovery by Heger in 1922 recalled recently in a nice form by
McCall and Griffin (2013). After more than 90 years they remain
unidentified; the identity of their carriers is thus the longest
standing unanswered question in all of spectroscopy. The current
list of unidentified interstellar absorption features, observable
in translucent interstellar clouds, contains more than 400 entries
(Hobbs et al., 2008, 2009). Despite many laboratory based studies
of possible DIB carriers, it has not been possible to
unambiguously link these bands to specific species. This is
unfortunate, as an identification of DIBs would substantially
contribute to our understanding of chemical processes in the
diffuse interstellar medium. The presence of substructures inside
DIB profiles, discovered by Sarre et al. (1995) and by Kerr et al.
(1998), supports the idea that DIBs are very likely molecular
features of numerous carriers that remain in the gas phase in
interstellar translucent clouds.

Establishing the rest wavelengths of DIBs is not a trivial task as
they have never been observed in laboratories. The only way to
accomplish this job is to move the whole spectrum to the rest
wavelength velocity frame using identified interstellar atomic or
molecular lines. The existing surveys display slightly different
central wavelengths for the same DIBs (Tab. \ref{waves}); this is
natural since the DIB profiles are nonsymmetric and thus the
measurements of their wavelengths depend on the chosen central
points. Moreover, in some cases DIB profiles are likely suffering
the Doppler split (Herbig \& Soderblom 1982; Weselak et al. 2010)
which makes the task of determining central wavelengths even more
complicated. To avoid uncertainties introduced by these problems
we have selected targets without an evident Doppler effect in
K{\sc i} atomic and CH molecular lines and have measured the displacement
of diffuse bands by cross-correlation with the template spectrum
from the survey by Galazutdinov et al. (2000) where measurements
are based on a single-cloud object HD23180. However, in almost
all cases the reddened OB stars are observed through more than one
cloud; i.e. the high resolving power is of critical importance. In
fact the DIB profiles are much broader than those of
atomic/molecular lines and thus their shapes are quite resistant
against the Doppler split and are not expected when the
split is not observed in atomic or molecular features.
Nevertheless, in a vast majority of cases we observe some
ill--defined averages along the chosen sightlines which makes the
determination of physical conditions, leading to the formation of
DIB carriers very complex.

Diffuse bands' carriers are believed to be concentrated in neutral
interstellar clouds which are revealed by K{\sc i} 7699~\AA\ line
(Table 1) or the methylidyne (CH) features (Megier et al. 2005).
The review by Galazutdinov et al. (2000) is based on the
assumption that the correction of the wavelength scale to the
rest-wavelength frame by means of either K{\sc i} or CH lines
provides also the correction of DIBs to their laboratory
wavelengths. The assumption seems justified in a vast majority of
cases (see, e.g. DIB surveys by Hobbs et al. (2008); Bondar
(2012)), but some exceptions have already been reported. Other
interstellar absorption lines cannot be used for this purpose;
e.g. the H and K lines of Ca{\sc ii} are apparently formed in
numerous tiny, Doppler shifted clouds which are mostly ionized
(Megier et al. 2009).

Kre{\l}owski \& Greenberg (1999) observed for the first time that
diffuse bands are red-shifted while related to the interstellar
Na{\sc i} lines in stars  HD37020, HD37022, HD37023,
HD37042, HD37061 belonging to the Ori OB1 association. The
observed red-shift was found to be quite small ($\sim$0.1~\AA) yet very evident
in all studied directions.
 The extinction curves of lines of sight to Ori OB1 targets are extremely peculiar
(see Figs. 4.8 and 4.9 in Fitzpatrick \& Massa, 2007) as well as interstellar spectrum: very weak molecular and
narrow diffuse bands while broader diffuse bands are well seen.
The targets in Ori OB1 are characterized by relatively weak simple
interstellar radicals CH, CN and narrow DIBs (with respect to the
color excess) making study of the positional displacement very
difficult. Interestingly, the weakening does not concern so called
"broad" diffuse bands, e.g. at  5780 and 6284~\AA.
So far the observed DIB's red-shifts are reported only in the
above mentioned stellar aggregate (Ori OB1).
Likely the peculiar physical conditions  due to e.g., strong UV-flux irradiation
from the nearby very hot young stars, do cause the observed displacement of diffuse bands.

The first object where some DIBs were reported as blue-shifted in
relation to either K{\sc i} or CH lines is HD34078 (Galazutdinov
et al. 2006). The paper also demonstrated that the profiles of
blue-shifted DIBs are broader than those observed in the spectrum
of HD23180 used for comparison. The widths of the narrow DIB
profiles seemingly grow towards blue. This moves the
centers of these features to the same direction (Fig. 1)
contributing to the net blue-shift effect.

According to Herbig (1999) the radial velocity of the CH 4300~\AA\
lines observed in HD34078 has changed between 1949 and 1999 from
13.6 to 15.3 km/s. This may coincide with the variable
intensity of this feature postulated by Rollinde et al. (2003).
However, our current measurement (15.4 km/s) coincides
with the 1999's value. It may suggest that the positional variability (if any)
does not progress in the linear way.

The phenomenon of DIB blue-shifts is typical for the Sco OB1
association. Galazutdinov et al. (2008) reported the detection of
blue-shifts for 20 diffuse bands, though some of the studied
features  are very weak and thus, the conclusions that concern
them are uncertain.

In this paper we present a study of positional displacement and
profile changes of diffuse bands in spectra of 4 objects (Tab.2,
3) including the previously studied HD34078 (Galazutdinov et al., 2006)
and HD152233 (Galazutdinov et al., 2008) but using another instrument.
We also observed two heavily reddened but apparently free of Doppler split effect objects HD73882
and HD169454.  We confirm the earlier results and extend the
number of studied diffuse bands using another instrument
(Magellan/Mike), now with much higher signal-to-noise ratio of the
carried out spectra. Thus, we confirm blue-shifts of many diffuse
bands in the spectrum of HD152323 as well as our former
detections for HD34078.  HD73882 and HD169454 in general do not reveal noticeable
displacements of diffuse bands.

\section{Observational material}

The spectra of HD34078 (AE Aur) and HD152233 (a member of Sco
OB1 association) were recorded together with those of HD73882
and HD169454 with the aid of the Magellan/Clay telescope at the
Las Campanas Observatory in Chile using the MIKE spectrograph
(Bernstein et al. 2003) with a 0.35$\times$5 arcs slit. The
recorded spectra are averages (in the pixel space) of 19 -- 60 individual exposures to achieve a
high S/N ratio.  We estimated the resolving power using the
solitary Thorium lines.
 It is $\sim$56,000 ($\Delta\lambda$ $\sim$5.4 km\,s$^{-1}$) on the blue
branch (3600-5000 \AA\AA) and $\sim$77,000 ($\Delta\lambda$ $\sim$3.9 km\,s$^{-1}$) on the red branch (4800-9400
\AA\AA). In all cases we used HD116658 (Spica) as a telluric line divisor.

 General information about observed lines of sight is given in Table 2, including the column density of neutral atomic and molecular
hydrogen.
The spectra were processed and measured in a standard way using
both IRAF (Tody 1986) and our own
DECH\footnote[1]{http://gazinur.com/DECH-software.html} codes.

\subsection{The wavelength calibration}

The wavelength scale of the studied spectra was calculated on the
basis of a global polynomial equation:

\begin{equation}
   \lambda(x,m) = \sum_{i=0}^{k} \sum_{j=0}^{n} a_{ij} x^{i} m^{j}
\end{equation}

where $a_{ij}$ are polynomial coefficients; $x$ - the pixel
position in the direction of dispersion;  m -- the order number.
The  final  solution  uses  typically  300-1200  lines  of Thorium
(depending on the spectrograph's arm) and the rms residual error
between the fit and the position of the lines is usually
$\leq$0.003~{\AA}, i.e. much lower than 1 km\,s$^{-1}$.

Prior to the measurements of displacement of diffuse bands, the wavelength scale
of all spectra was shifted to move the CH 4300~\AA\ line to their
laboratory positions, i.e. the rest wavelength velocity frame was
set  interstellar.  Thereto, the wavelength correction in Angstroms was calculated for each
pixel of the spectrum based on the measured CH 4300 shift in the radial velocity (km\,s$^{-1}$) scale.

The selection of CH as a rest point is argued by
the fact that in most of surveys the radial velocity of the cores
of profiles of CH 4300 and KI 7699 are almost identical. Moreover,
ultra-high resolution profiles of CH 4300 and K{\sc i} 7699 in spectra
of HD23180  studied by Crane, Lambert \& Sheffer (1995) and
Welty \& Hobbs (2001) report the radial velocity values of the
cores of features as +14.6 and +13.5 km\,s$^{-1}$ respectively,
i.e. the wavelength uncertainty does not exceed 0.01~\AA.

\subsection{Full width at the half of the maximum (FWHM) of the diffuse bands}

Width of diffuse bands cannot be precisely measured by Gaussian or Voigt fit because
of irregular shape of profiles. To measure the FWHM, the studied profiles were smoothed by a Fourier filter.
Then, the minimum of the smoothed curve is found as well as
two required points of the smooth curve at the half-intensity. The method was well illustrated by Bondar (2012) in his figure 3
with a difference that Bondar used a Chebyshev polynomial fit instead of our Fourier smoothing applied.

\subsection{Measurements of displacement}

To minimize the effects of uncertain rest wavelengths we measured
the displacement of diffuse bands by the cross-correlation of
profiles of diffuse bands in the studied objects with those of the
well-studied line of sight HD23180 - an object with the
negligible Doppler split.  Id est, diffuse bands profiles
of each studied spectrum were cross-correlated with the
corresponding profiles of the template. Then, the measured
displacement was applied to the rest wavelength of the studied
diffuse band. We might note again, that these rest wavelengths
were determined as means of that particular object HD23180, see
our survey Galazutdinov et al., 2000.

For most of diffuse bands we have used as a template the average spectrum of HD23180 from our survey
(Galazutdinov et al. (2000)) and, in cases of weaker diffuse bands, we
used more recent spectra of the same target from our database.
Before the measurements, the template spectrum profiles were smoothed using the Fourier
transform based filter. Measurements of profile displacements have
been done using our DECH code procedure based on the algorithm
developed by Tonry \& Davis (1979) and Verschueren \& David (1999).

 Cross-correlation of two sufficiently symmetric (e.g., DIBs 5780 and 5850, see Fig. 2)
profiles produces also symmetric Gaussian-like cross correlation function (CCF), where the position of the maximum is evident (Fig.3).
However, most of the diffuse bands do not exhibit symmetric profiles (e.g. DIB 5797). Nevertheless, the close similarity
of measured profiles with those of the template also produce sufficiently symmetric CCF with
evident position of the maximum of the correlation.

Some diffuse bands are broad and shallow making their analysis rather difficult (e.g. DIB 5776): continuum normalization plays a
crucial role in formation of the shape of subsequent cross-correlation functions (CCF), although does not affect much to the peak position,
which can be estimated without a doubt.

 DECH software continuum normalization method is based on the cubic spline interpolation {\it over} the manually set
fiducial marks.  As it was mentioned above, the continuum normalization plays a crucial role for the shape of
the subsequent CCF. Even in complicated cases (e.g., noisy and weak DIB with irregular asymmetric
profile) the resulting CCF enable the confident finding the top position, i.e. the value of displacement.

Results of measurements are presented in Tab. \ref{measdata} and Fig.4.

\section{Results}

In the present survey we consider a sample of diffuse bands free
of evident blending with stellar lines and being strong enough to
provide sufficiently high S/N ratios inside their profiles. The
four considered spectra were shifted to the rest wavelength
velocity scale using the CH 4300~\AA. In Fig. 5  CH 4300~\AA\ line
is shown together with CH B-X (0,0) and CN B-X (0,0) bands in the
blue range. The spectra in the left panel of Fig. 5 are normalized
to identical depths of the CH 3886~\AA\ feature. It is evident
that the position of the lines carried by simple interstellar
radicals coincide in all targets in the frame of measurement
uncertainties. This suggests that all possible displacements of
interstellar features are likely caused by physical modifications
of DIB profiles unless some clouds, not revealed by the depicted
lines, are present along the sightlines. The relative (to CH)
abundance of CN is much higher (by an order of magnitude) towards
our comparison objects: HD73882 and HD169454 (see also
Kre{\l}owski et al. 2012). This considerable difference in the
relative abundances of simple interstellar radicals proves
different physical/chemical conditions in the clouds situated
along the sight lines to our targets.

In all our targets we can trace the C$_3$ \~{A} $^{a}\Pi_{u}$ -
\~{X} $^{a}\Sigma_{g}^{+}$ 000-000 band (Fig. 6). The point to be
emphasized is that the rotational temperature of HD34078 of
C$_3$ molecule is very high in contrast to HD169454 --  see Fig.
6 confirming results of {\'A}d{\'a}mkovics et al. (2003).
According to them T$_{rot}$ equals 42K for HD169454 and 171K for
HD34078. {\bf Our recent result for C$_3$ T$_{rot}$ is 21$\pm$1 K.}
T$_{rot}$ observed towards our standard, HD23180 is
90K, i.e. resides between HD169454 and HD34078.

We have measured  positional displacement, FWHM (full widths at the half
maxima) and equivalent widths of 29 diffuse bands (Table
\ref{measdata} and Fig. 4). Our measurements confirm the
blue-shifts of the DIBs reported by Galazutdinov et al. (2006) in
the spectrum of HD34078. Similar effect is seen towards
HD152223 (previously reported by Galazutdinov et al. (2008)) --
some bands show even greater displacements than those in
HD34078. Evidently these two targets show a "blue-shift" of most
of diffuse bands in general.
Generally speaking, careful measurements reveal a common
``instability'' of positions of diffuse bands from object to
object with amplitude within $\sim$ $\pm$5km\,s$^{-1}$.

The main DIBs (the first ever observed) near 5780 and 5797~\AA\
are shown in Fig. 2 along with their weaker neighbors 5776 and
5850. Both  DIB5797 and DIB5776 are evidently blue-shifted in
the spectra of HD34078 and in HD152233 confirming the result
of Galazutdinov et al. (2008). The 5797 profile is substantially
broadened in the spectrum of HD34078 in comparison to HD73882
but, the 5797 profile in HD152233, though also blue--shifted, is
not broadened at all (Fig. 1). Evident broadening of many diffuse
bands is well seen in the spectra of HD34078 (Fig.7a and 7b).
Interestingly, in the case of HD152233 the evident blue shift
seen in many diffuse bands (Fig. 4) is not accompanied with the
broadening of their profiles (Fig. 7a).

 It is worth noting that one of the major diffuse bands, strong and broad DIB5780
exhibits {\bf displacement} of the peak position: indeed, while the feature is slightly blue
shifted in spectra of HD34078 and HD152233, the evident red-shift of the DIB is seen in
the spectrum of HD169454 where the diffuse band is strongest in the sample.
It may be interesting to reveal possible relation between the positional displacement
and strength (i.e. reddening) of this band.

The notable positional displacement can be observed in the
case of the narrow 6196 DIB (Galazutdinov et al. 2006, 2008). The variable profile of this DIB
was related to the $T_{rot}$ of the C$_2$ molecule by Ka\'zmierczak et al. (2009).

Let's discuss finally the possible reasons for the observed
wavelength shifts of diffuse bands using the most narrow and
relatively deep DIB at 6196, demonstrating evident displacements.
The estimates of its central wavelength are a bit floating
(Table \ref{waves}) from author to author, but in general they
remain in the range of 0.05~\AA. Fig.8 presents DIB6196 shown
together with four identified interstellar lines: CH -- 4300~\AA,
CH$^+$ -- 4232~\AA, Ca{\sc ii} -- 3933~\AA\ and K{\sc i} --
7699~\AA\ in the spectra of our targets in the scale of
heliocentric radial velocities.

The comparison suggests the following conclusions:
\begin{itemize}
\item
radial velocities of K{\sc i} and CH lines are always very close
\item
the radial velocity of the 6196 DIB does not always agree with
those of K{\sc i}, CH and CH$^+$
\item
the position of 6196 in HD34078 may coincide with a very weak
Doppler component of K{\sc i}, having no correspondence in profiles of
molecular lines;
\item
in the case of HD152233 the DIB position coincides with weaker
(but still strong) component of CH$^+$ -- this component is very
weak in the CH profile, but present in Ca{\sc ii} and Na{\sc i}
lines.

\end{itemize}

\section{Discussion and Conclusions}

The published conclusions of Galazutdinov et al. (2006, 2008) are
now confirmed using an additional instrument from another
observatory. Moreover we extended our considerations to the weaker
DIBs which were out of reach while using spectra from the Terskol
and Bohyunsan observatories.  The average displacement of 29
measured diffuse bands relative to CH 4300 \AA\ line is
+1.0$\pm$2.1 km\,s$^{-1}$ for HD73882, +0.8$\pm$2.9 km\,s$^{-1}$
for HD169454, -4.0$\pm$3.8 km\,s$^{-1}$ for HD34078 and
-3.6$\pm$3.1 km\,s$^{-1}$ for HD152233.
 However, just 18 out of measured 29 diffuse bands exhibit
blue shift in both directions to HD152233 and HD34078. Two weak diffuse bands
in blue range (DIBs 4985 and 5542) are blue-shifted in HD152233 only. Also,
there are weak 3 bands (6117, 6140 and 6376) with measured blue-shift in HD34078
but not firmly confirmed in HD34078.

Our Fig.7 proves the general broadening of diffuse band in HD34078. However,
the relation between the broadening magnitude and displacement is not revealed: see Fig. 7b.

The diffuse bands wavelength variations are thus proven beyond a
doubt, though the origin of the shifts remains unclear. Our sample
of targets is evidently too small to make a definite conclusion
that DIB positions are related to the $T_{rot}$ of e.g. simple
carbon chains since such a conclusion would be based on three
points only. HD34078 is a very peculiar object in which the
C$_3$ rotational temperature is the highest ever observed
({\'A}d{\'a}mkovics et al., 2003) but HD152233 seems to be very
similar in this respect. Our standard, HD23180, represents an
average $T_{rot}$ and, usually, also an average DIB position.
Temperature variations, acting more or less like in the profile of
C$_3$ band but in species of lower rotational constants, may
change the observed DIB profiles, shifting their centers
to the blue and altering the substructure systems seen in DIB
profiles (see e.g. Fig.11 in Motylewski et al. (2000)).  The most abundant interstellar molecule $H_2$ was also
proven to be at highly excited rotational states towards HD34078
(Boisse et al. 2005). Moreover, the CH abundance in its direction
is twice as big as in average for such an E(B-V); HD152233 does
not share this property. A variability of the CH and CH$^+$
abundance in HD34078 was already suggested by Boisse et
al.(2009).

The observed interstellar features, seen in the spectra of
HD34078 and HD152233 are evidently originating in specific
environments and deserve more observation. However, the behavior
of DIBs in both objects is not identical. As shown in Fig. 2 one
of main DIBs (5797) is blue-shifted in the spectra of the above
targets if compared to that of HD73882. However, if to normalize
the 5797 DIB to the same depth and to remove the shift in
HD152233 (for clarity), one can see that the profile shape of
this feature is identical in the latter object and in HD73882. On
the other hand the DIB profile in HD34078 is much broader and only
its center of gravity moves a bit blue-ward. Thus what we observe
in HD152233 may be just a result of the Doppler shift, if the
DIB originates in the cloud revealed by the CH$^+$ component(see
also how the 6196 DIB behaves in HD152233 (Fig. 8) while in
HD34078 it likely follows very specific environmental
conditions.

Thus, we can summarize the possible explanations of observed
displacements of diffuse bands:
\begin{itemize}
\item
the peculiar environmental conditions revealed by e.g. high
rotational temperatures of simple molecules; a good example
is HD34078 where we observe unusually high rotational
temperatures of simple molecular species, such as C$_2$, C$_3$ or
CN. We believe that CH and CH$^+$ can be rotationally excited
but our existing spectra show too low S/N ratio -- insufficient for
checking such subtle effects. We have already demonstrated
(Galazutdinov et al. 1998) that DIBs can be observed when color
excess is practically zero. One can expect such components in
certain DIBs originated in peculiar conditions.
\item
simply due to Doppler effect, i.e. unresolved multiple Doppler
components that cause the movement of the center of the broadened
profile (see Fig. 8); but in this case one meets a cloud with
peculiar chemical composition or unusual physical parameters, i.e
the "shifted" diffuse band does not correspond to the main (deepest)
CH or K{\sc i} component (Fig. 8), i.e. the diffuse band carrier(s)
are situated somewhere apart from the main "body" of the cloud.
\end{itemize}

More definite result requires much more representative sample of detected DIB shifts
though one important conclusion can be inferred: despite of good general correlation between
diffuse bands and interstellar KI and CH (normally serving for "interstellar" wavelength scale correction),
diffuse bands carriers may not be spatially correlated with simple atomic/molecular gas.

\begin{acknowledgements}

GG acknowledges the support of Chilean fund FONDECYT-regular
(project 1120190). JK acknowledges the financial support of the
Polish National Center for Science during the period 2012 - 2015
(grant UMO-2011/01/BST2/05399). GV acknowledges
the support by the Russian Scientific Foundation (grant N 14-50-00043
We are grateful for the assistance of the Las Campanas observatory staff members.

\end{acknowledgements}

\clearpage

\begin{table}
\caption{Comparison of the rest wavelength of diffuse bands at 6196 and 6614 \AA\ in different surveys. "Line" marks the features used to move
the spectra to the rest wavelength frame.}
\begin{tabular}{lcll}
\hline
     Survey                  & line                   &  DIB 6196  & DIB 6614  \\
\hline
Galazutdinov et al. (2000)   & K{\sc i} 7699          &  6195.97      & 6613.56  \\
Weselak et al. (2000)        & Na{\sc i} D$_1$, D$_2$ &  6196.05      & 6613.52  \\
Hobbs et al. (2008)          & K{\sc i} 7699          &  6195.98      & 6613.62  \\
Hobbs et al. (2009)          & K{\sc i} 7699          &  6196.09      & 6613.70  \\
Bondar (2012)                & CH 4300                &  6195.95      & 6613.58 \\
\hline
\end{tabular}
\label{waves}
\end{table}

\begin{table}
\caption{Observed targets. Column densities of molecular and neutral hydrogen are from Gudennavar et al. (2012)
Rotational temperature for C$_3$ and C$_2$ molecules (lower transitions) are taken from: a -  {\'A}d{\'a}mkovics et al. (2003); b - Schmidt et al. (2014); c - Sonnentrucker et al. (2007)    .
}
\scalebox{0.8}{
\begin{tabular}{rrlclllll}
\hline
     HD     & V   &  Sp/L      &  E(B-V)  & N(H I) $\times$10$^{21}$              &  N(H$_2$) $\times$10$^{20}$    & H$_2$ fraction & C$_3$ T$_{rot}$ (K) & C$_2$ T$_{rot}$ (K)$^{a}$  \\
\hline
   34078    & 6.0 &  O9.5V     &  0.53    & 1.5849$^{+0.4568}_{-0.3546}$          &  7.5858                        &     0.32       &  171$^{a}$          &  120$\pm$10$^{a}$  \\
   73882    & 7.3 &  O9III     &  0.70    & 1.2882                                & 12.882$^{+2.6057}_{-2.1673}$   &     0.5        &                     &  23$\pm$5$^{c}$  \\
  152233    & 6.6 &  O6III:(f)p&  0.45    & 1.9498                                &  1.9498                        &     0.1        &                     &   \\
  169454    & 6.7 &  O+...     &  1.12    & 0.08912                               & 14.454                         &     0.94       &  21$\pm$1$^{b}$     & 50$\pm$10$^{a}$   \\
\hline

\end{tabular}
}
\label{stars}
\end{table}

\begin{table}
\caption{The measured equivalent widths (EW in m\AA),
 positional displacement (kmº,s$^{-1}$) relative to CH 4300,  and full-width at the half-maximum (RV and FWHM  in km\,s$^{-1}$)  of the interstellar
features relative to the rest wavelength position of CH 4300.313 \AA\ line.
Wavelength data source: g - Galazutdinov et al. (2000)  measured in HD23180; h - Hobbs (2008)  measured in HD204827.
The last column exhibits the displacement span (in km\,s${-1}$).}
\scalebox{0.6}{
\begin{tabular}{l|rrrcrrrcrrrcrrrr}
\hline
              &  \multicolumn{3}{c}{HD73882}                 & & \multicolumn{3}{c}{HD169454}                & &  \multicolumn{3}{c}{HD34078}             &   &  \multicolumn{3}{c}{HD152233}    \\
\cline{2-4}
\cline{6-8}
\cline{10-12}
\cline{14-16}
\\
              & \multicolumn{1}{c}{RV}
                                 &\multicolumn{1}{c}{EW}
                                                & \multicolumn{1}{c}{FWHM}
                                                             & &
                                                                 \multicolumn{1}{c}{RV}
                                                                                 & \multicolumn{1}{c}{EW}
                                                                                                & \multicolumn{1}{c}{FWHM}
                                                                                                             & &
                                                                                                                \multicolumn{1}{c}{RV}
                                                                                                                             & \multicolumn{1}{c}{EW}
                                                                                                                                             & \multicolumn{1}{c}{FWHM}
                                                                                                                                                          & &
                                                                                                                                                            \multicolumn{1}{c}{RV}
                                                                                                                                                                          & \multicolumn{1}{c}{EW}
                                                                                                                                                                                           & \multicolumn{1}{c}{FWHM}
                                                                                                                                                                                                            & \multicolumn{1}{r}{span}\\
\hline
FeI 3859.9114 &     -3.6$\pm$0.1 &  3.9$\pm$0.4 &  15$\pm$3: & &     -1.1$\pm$0.5&   5.0$\pm$1.5& 10$\pm$5:  & & +0.2$\pm$0.2&   6.4$\pm$0.1 & 8.9$\pm$0.2& & -2.0$\pm$0.1&    3.0$\pm$0.2 & 9.0:$\pm$0.3  & 3.8  \\
CN  3873.994  &     +0.3$\pm$0.1 & 19.8$\pm$0.2 & 6.1$\pm$0.2& &     -0.1$\pm$0.2&  16.7$\pm$0.2& 5.6$\pm$0.2& & -0.7$\pm$0.5&   2.9$\pm$0.4 & 8.1$\pm$0.1& & +0.5$\pm$0.2&    0.5$\pm$0.2 & 6.0:$\pm$1.0  & 1.2  \\
CN  3874.602  &     +0.3$\pm$0.1 & 34.6$\pm$0.3 & 6.3$\pm$0.2& &     -0.2$\pm$0.2&  23.3$\pm$0.3& 5.8$\pm$0.2& & -0.6$\pm$0.2&   9.0$\pm$2.0 & 8.5$\pm$0.5& & -0.3$\pm$0.1&    1.7$\pm$0.3 &  6.8$\pm$0.2  & 0.9  \\
CN  3875.759  &     +0.4$\pm$0.1 & 12.0$\pm$0.3 & 6.0$\pm$0.2& &     -0.2$\pm$0.1&  11.5$\pm$0.2& 5.6$\pm$0.2& & -1.5$\pm$0.7&   1.5$\pm$1.0 & 9.1$\pm$0.2& & +0.1$\pm$0.3&    0.3$\pm$0.2 &  6.5$\pm$0.5  & 1.9  \\
CH  3886.409  &     +0.0$\pm$0.1 &  7.7$\pm$0.3 & 7.8$\pm$0.5& &     -0.4$\pm$0.2&   7.9$\pm$0.4& 6.5$\pm$0.1& & -0.6$\pm$0.1&  19.5$\pm$1.0 & 8.9$\pm$0.2& & -0.3$\pm$0.1&    3.4$\pm$0.2 &  8.4$\pm$0.3  & 0.6  \\
CH  3890.217  &     +0.0$\pm$0.1 &  5.7$\pm$0.3 & 7.3$\pm$0.5& &     -0.5$\pm$0.2&   5.6$\pm$0.4& 6.5$\pm$0.1& & -0.6$\pm$0.5&  10.9$\pm$0.7 & 9.1$\pm$0.2& & -0.2$\pm$0.1&    2.0$\pm$0.2 &  8.2$\pm$0.2  & 0.6  \\
CaI 4226.728  &     -4.9$\pm$0.1 & 57.5$\pm$0.5 & 7.1$\pm$0.2& &     +0.5$\pm$0.3&  16.3$\pm$0.9&10.5$\pm$0.3& & +3.0$\pm$1.0&   6.9$\pm$0.6 &14.5$\pm$0.5& & +0.0$\pm$0.2&    7.6$\pm$0.4 &  6.6$\pm$0.1  & 7.9  \\
CH+ 4232.548  &     -0.2$\pm$0.1 & 18.4$\pm$0.4 & 8.1$\pm$0.1& &     -0.8$\pm$0.3&  17.9$\pm$0.9& 8.5$\pm$0.3& & -0.2$\pm$0.2&  38.3$\pm$0.5 & 8.5$\pm$0.2& & -0.8$\pm$0.2&   21.6$\pm$0.2 & 14.0$\pm$0.1  & 0.6  \\
CH  4300.313  &     +0.0$\pm$0.1 & 24.6$\pm$0.3 & 6.5$\pm$0.1& &     +0.0$\pm$0.1&  28.5$\pm$0.5& 6.4$\pm$0.2& & +0.0$\pm$0.1&  50.1$\pm$0.5 & 7.8$\pm$0.1& & +0.0$\pm$0.1&   12.8$\pm$0.2 &  6.9$\pm$0.1  & 0.0  \\
KI  7698.965  &     -0.4$\pm$0.2 & 48.0$\pm$0.8 & 8.7$\pm$0.3& &     -0.4$\pm$0.2&   211$\pm$1.0& 8.1$\pm$0.2& & +0.0$\pm$0.1& 171.0$\pm$2.0 & 9.1$\pm$0.3& & +0.0$\pm$0.1&  127.0$\pm$1.0 &  6.7$\pm$0.2  & 0.4  \\
\hline
DIB4726.33g   &     +0.6$\pm$2.3 & 56.0$\pm$9.9 & 190$\pm$9  & &     -0.9$\pm$4.3& 140.0$\pm$50.& 180$\pm$50 & & +0.1$\pm$1.5&  99.3$\pm$9.9 &190$\pm$5   & & -1.8$\pm$3.5&  67.0$\pm$5.0  &  190$\pm$5    & 2.4 \\
DIB4734.79h   &     +5.5$\pm$2.0 &  1.9$\pm$0.4 &  20$\pm$5  & &     +2.5$\pm$1.0&   6.2$\pm$1.0&  31$\pm$5  & & +4.5$\pm$1.5&  14.3$\pm$4.0 & 42$\pm$2   & & +3.0$\pm$2.1&   4.2$\pm$0.6  &   26$\pm$3    & 3 \\ 
DIB4762.52g   &     -2.0$\pm$2.4 & 43.5$\pm$4.0 & 140$\pm$5  & &     -1.0$\pm$3.0& 115.0$\pm$15.& 145$\pm$10 & & -5.5$\pm$1.6&  93.5$\pm$9.0 &135$\pm$10  & &-12.0$\pm$3.0&  50.0$\pm$9.0  &  140$\pm$20   & 11 \\ 
DIB4963.85g   &     -1.5$\pm$0.5 &  9.2$\pm$1.5 &  43$\pm$3  & &     -2.0$\pm$1.0&  26.3$\pm$1.2&  41$\pm$2  & & -4.4$\pm$1.6&  20.0$\pm$2.0 & 48$\pm$1   & & -3.6$\pm$0.3&  12.4$\pm$0.6  &   39$\pm$1    & 2.9 \\ 
DIB4984.77g   &     -2.5$\pm$3.5 &  4.9$\pm$1.3 &  35$\pm$2  & &     -5.5$\pm$1.0&  14.1$\pm$1.3&  31$\pm$1  & & -9.5$\pm$2.0&   7.9$\pm$1.0 & 37$\pm$1   & & -6.0$\pm$4.5&   6.0$\pm$0.4  &   31$\pm$2    & 7 \\ 
DIB5494.07g   &     +1.0$\pm$2.0 &  9.6$\pm$1.5 &  37$\pm$5  & &     -1.2$\pm$0.2&  23.4$\pm$1.6&  31$\pm$2  & & -6.0$\pm$1.0&  10.2$\pm$1.2 & 31$\pm$4   & & -6.0$\pm$1.0&  12.8$\pm$1.2  &   35$\pm$2    & 7 \\
DIB5512.65g   &     +0.0$\pm$2.0 &  2.0$\pm$1.0 &  34$\pm$3  & &     -0.2$\pm$0.2&  15.3$\pm$1.3&  31$\pm$1  & & -1.2$\pm$0.4&   6.6$\pm$1.0 & 46$\pm$3   & & -1.5$\pm$1.3&   6.7$\pm$0.8  &   32$\pm$1    & 1.5 \\ 
DIB5541.74g   &     -1.5$\pm$0.7 &  2.2$\pm$1.2 &  39$\pm$4  & &     +1.3$\pm$0.4&   8.5$\pm$1.0&  32$\pm$2  & & +1.2$\pm$2.0&   6.2$\pm$0.8 & 35$\pm$3   & & -1.5$\pm$0.5&   5.2$\pm$0.9  &   41$\pm$1    & 2.8 \\ 
DIB5544.95g   &     +0.0$\pm$2.0 &  6.4$\pm$1.0 &  47$\pm$2  & &     +3.0$\pm$2.0&  25.7$\pm$1.2&  44$\pm$2  & & -6.3$\pm$2.2&   3.8$\pm$0.6 & 46$\pm$3:  & & -3.4$\pm$0.5&   8.9$\pm$1.5  &   44$\pm$2    & 9.3 \\
DIB5546.46g   &     -0.3$\pm$0.5 &  0.8$\pm$0.5 &  23$\pm$2  & &     -4.0$\pm$2.0&   7.3$\pm$1.5&  30$\pm$3  & & -4.5$\pm$1.0&   1.3$\pm$0.5 & 22$\pm$5   & & -3.5$\pm$1.7&   3.4$\pm$0.5  &   31$\pm$4    & 4.2 \\ 
DIB5775.89g   &     +0.5$\pm$1.0 &  5.7$\pm$1.1 &  50$\pm$5  & &     +2.0$\pm$0.7&  13.0$\pm$2.0&  58$\pm$2  & & -9.5$\pm$3.0&   7.0$\pm$2.0 & 43$\pm$1   & & -3.5$\pm$2.0&   5.0$\pm$1.3  &   62$\pm$5    & 11.5 \\ 
DIB5780.37g   &     +7.5$\pm$0.5 &145.0$\pm$5.0 & 106$\pm$5  & &     +8.0$\pm$1.0& 467.0$\pm$5.0& 110$\pm$1  & & +2.5$\pm$1.0& 165.0$\pm$5.0 &110$\pm$1   & & -1.7$\pm$0.3& 215.0$\pm$3.0  &  105$\pm$1    & 9.7 \\ 
DIB5796.99g   &     +0.5$\pm$1.8 & 34.0$\pm$1.4 &  40$\pm$3  & &     +4.0$\pm$0.8& 161.0$\pm$1.8&  44$\pm$1  & & -4.3$\pm$1.4&  56.0$\pm$1.5 & 51$\pm$2   & & -2.0$\pm$1.3&  60.0$\pm$2.8  &   43$\pm$1    & 8.3 \\ 
DIB5849.82g   &     -1.3$\pm$0.9 & 14.7$\pm$1.8 &  50$\pm$2  & &     +0.7$\pm$0.1&  68.6$\pm$1.8&  42$\pm$1  & & -2.0$\pm$1.8&  31.0$\pm$1.5 & 60$\pm$4   & & -3.0$\pm$0.5&  28.0$\pm$2.0  &   45$\pm$1    & 3.7 \\ 
DIB6065.31g   &     +3.3$\pm$3.0 &  2.5$\pm$0.5 &  30$\pm$7  & &     +4.2$\pm$1.0&   9.8$\pm$0.5&  29$\pm$3  & & -0.9$\pm$0.3&   6.8$\pm$1.0 & 30$\pm$3   & & -0.9$\pm$0.3&   2.5$\pm$0.4  &   27$\pm$1    & 5.1 \\ 
DIB6089.78g   &     +0.5$\pm$2.0 &  4.1$\pm$0.6 &  30$\pm$3  & &     +0.7$\pm$0.6&  26.3$\pm$0.6&  28$\pm$1  & & -7.0$\pm$1.0&   4.2$\pm$0.6 & 30$\pm$3   & & -2.0$\pm$1.0&   6.6$\pm$0.5  &   28$\pm$2    & 7.7 \\ 
DIB6108.05g   &     +1.3$\pm$2.0 &  2.5$\pm$0.5 &  20$\pm$3  & &     +0.0$\pm$0.5&   8.8$\pm$0.8&  24$\pm$3  & & -6.0$\pm$4.0&   4.6$\pm$0.9 & 36$\pm$3   & & -6.9$\pm$0.3&   3.4$\pm$0.3  &   26$\pm$1    & 8.2 \\
DIB6113.20g   &     +0.5$\pm$0.7 &  6.9$\pm$0.6 &  55$\pm$9  & &     -0.5$\pm$1.0&  18.5$\pm$1.7&  34$\pm$3  & & +0.5$\pm$1.0&   9.5$\pm$0.7 & 60$\pm$4   & & -2.5$\pm$2.7&  14.0$\pm$1.3  &   48$\pm$1    & 3 \\ 
DIB6116.84h   &     +1.3$\pm$0.3 &  5.0$\pm$1.5 &            & &     -2.0$\pm$4.0&  13.8$\pm$1.4&  44$\pm$1  & &-12.5$\pm$2.5&   9.6$\pm$1.1 & 64$\pm$2   & & -0.1$\pm$1.5&   5.6$\pm$1.0  &               & 13.8 \\
DIB6139.94g   &     -0.3$\pm$1.5 &  4.0$\pm$1.0 &  25$\pm$2  & &     +0.0$\pm$0.7&  13.0$\pm$1.0&  28$\pm$3  & & -6.0$\pm$1.0&   3.5$\pm$1.1 & 37$\pm$2   & & -0.5$\pm$1.0&   6.0$\pm$1.0  &   60$\pm$30   & 6 \\
DIB6161.84g   &     +0.7$\pm$1.0 &  1.2$\pm$0.5 & 30:$\pm$9  & &     +2.8$\pm$1.0&   8.2$\pm$0.5&  21$\pm$2  & & -2.7$\pm$1.0&   2.5$\pm$0.8 & 30$\pm$5   & & -5.6$\pm$1.0&   3.5$\pm$1.0  &   21$\pm$5    & 8.4 \\ 
DIB6195.97g   &     +3.7$\pm$0.5 & 16.6$\pm$0.5 &  23$\pm$3  & &     +1.8$\pm$0.2&  62.0$\pm$1.0&  24$\pm$2  & & -6.0$\pm$0.5&  22.0$\pm$1.0 & 26$\pm$5   & & -5.6$\pm$0.2&  24.0$\pm$1.0  &   24$\pm$3    & 9.7 \\ 
DIB6203.08g   &     +0.5$\pm$1.0 & 29.6$\pm$1.5 &  63$\pm$1  & &     +1.8$\pm$1.0&  82.8$\pm$3.0&  59$\pm$1  & & -5.2$\pm$3.0&  37.0$\pm$1.5 & 69$\pm$1   & & -9.5$\pm$1.0&  50.0$\pm$1.5  &   67$\pm$2    & 11.3 \\
DIB6269.75g   &     +1.5$\pm$2.0 & 33.5$\pm$3.5 &  57$\pm$2  & &     +1.7$\pm$0.5& 104.0$\pm$5.8&  57$\pm$2  & & -3.5$\pm$1.0&  64.4$\pm$3.9 & 67$\pm$3   & & -7.2$\pm$0.5&  44.5$\pm$5.0  &   60$\pm$4    & 8.9 \\
DIB6283.85g   &     +1.1$\pm$0.5 &              & 187$\pm$9  & &     +7.3$\pm$5.0&              & 173$\pm$3  & & -1.4$\pm$3.0&               &            & & -0.7$\pm$1.9&                &  170$\pm$3    & 8.7 \\
DIB6375.98g   &     +0.0$\pm$1.5 & 10.2$\pm$1.5 &            & &     +2.0$\pm$2.0&  51.0$\pm$3.5&  38$\pm$4  & & -4.0$\pm$2.5&  15.0$\pm$1.7 & 66$\pm$5   & & -1.0$\pm$1.5&  23.2$\pm$2.2  &   55$\pm$5    & 6 \\ 
DIB6379.32h   &     +3.1$\pm$0.2 & 12.3$\pm$0.7 &  31$\pm$2  & &     +0.3$\pm$0.1&  93.6$\pm$3.0&  28$\pm$1  & & -6.0$\pm$0.9& 188.0$\pm$2.0 & 35$\pm$3   & & -2.7$\pm$0.2&  24.6$\pm$1.7  &   29$\pm$2    & 9.1 \\ 
DIB6613.56g   &     +2.0$\pm$1.0 & 46.0$\pm$1.1 &  47$\pm$1  & &     -0.5$\pm$0.5& 192.0$\pm$3.0&  45$\pm$1  & & -5.0$\pm$1.5&  63.0$\pm$2.0 & 55$\pm$2   & & -5.0$\pm$0.5&  82.0$\pm$2.0  &   46$\pm$2    & 7 \\
DIB6660.64g   &     +2.0$\pm$0.5 &  5.7$\pm$0.7 &  29$\pm$1  & &     -2.3$\pm$0.3&  34.4$\pm$0.9&  26$\pm$2  & & -8.5$\pm$1.0&   6.0$\pm$2.0 & 28$\pm$5   & & -7.0$\pm$1.5&  11.7$\pm$1.4  &   27$\pm$3    & 10.5 \\  
\hline
\end{tabular}
}
\label{measdata}
\end{table}

\clearpage

\begin{figure}
\center{\includegraphics[angle=270,width=90mm]{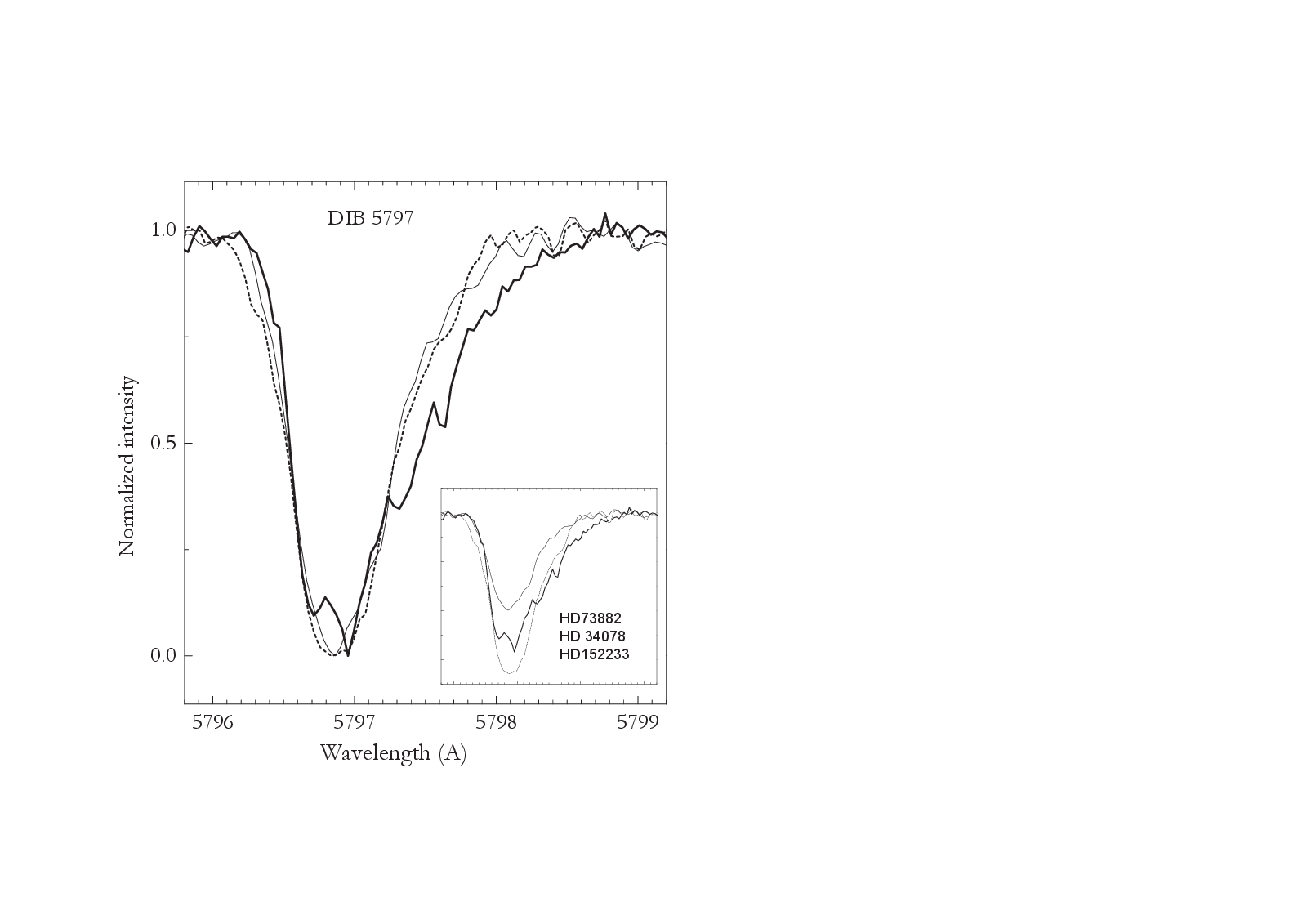}
 \caption{
   Profile of DIB 5797 in spectra of 3 program stars normalized to the same depth. Original profiles are shown in the small panel.
   The absolute wavelength scale is arbitrary -- all profiles are shifted for more evident comparison of profiles.
   Thick solid line - HD34078, thin solid line - HD73882, dash line - HD152233.
   FWHM of the feature is almost identical in HD152233 and HD73882, i.e. the blue shift observed in HD152233 may suggest a Doppler origin
   while that of HD34078 can be due to the profile broadening.
 }
}
\label{broaden}
\end{figure}

\begin{figure}
\center{\includegraphics[angle=270,width=15cm]{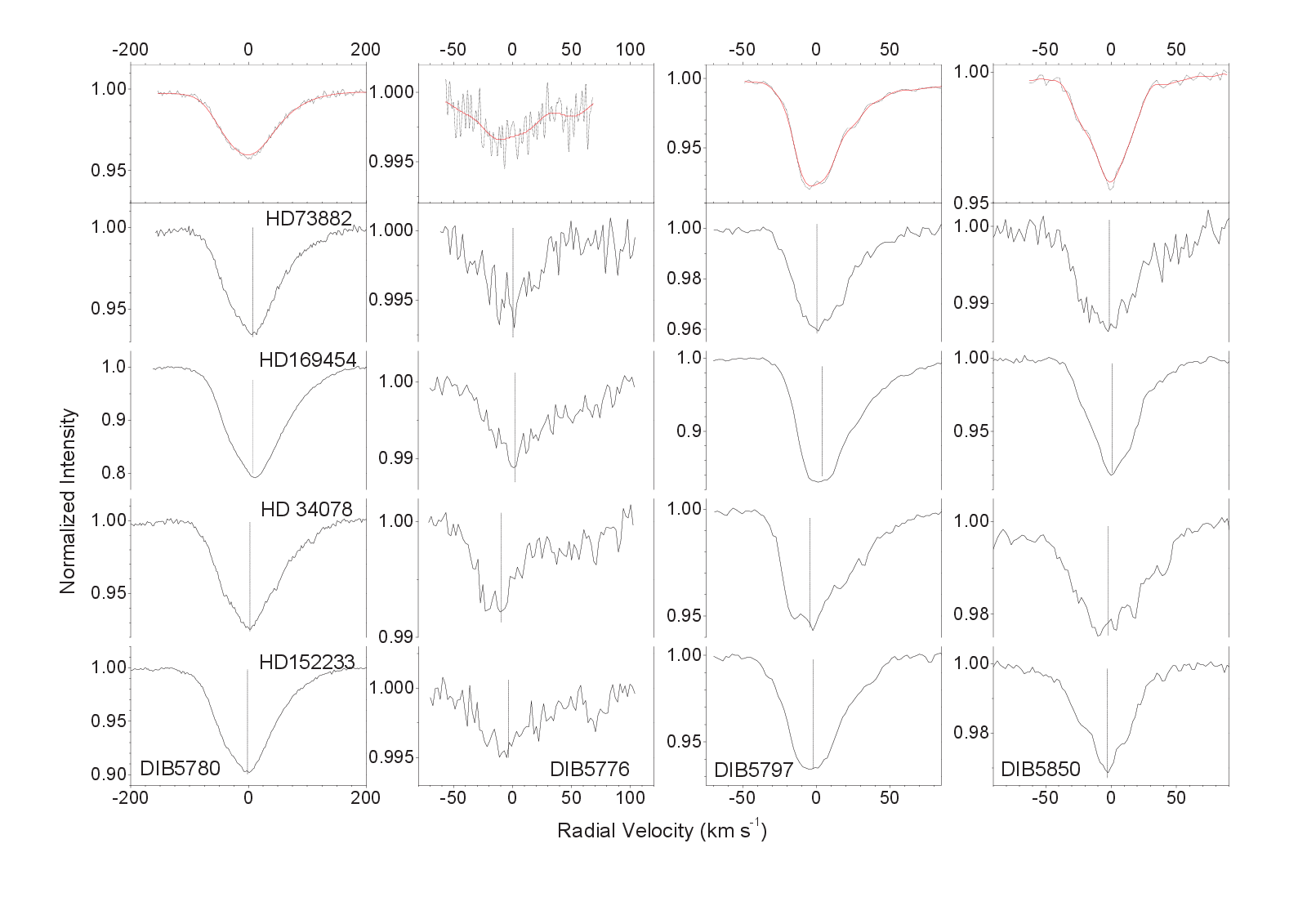}
  \caption{
  A sample of diffuse bands in studied spectra (bottom) shown together with a template spectrum (top). The template spectrum (the upper panels) is shown in its original form (black dash line)
  and as the Fourier-smoothed profiles (red solid lines). The latter were cross-correlated with studied DIBs for measurement the positional displacement. The vertical lines
  correspond to the  measured shift position.
 (A color version of this figure is available in the online journal.)
  }
}
\label{comp5780}
\end{figure}

\begin{figure}
\center{\includegraphics[angle=270,width=8cm]{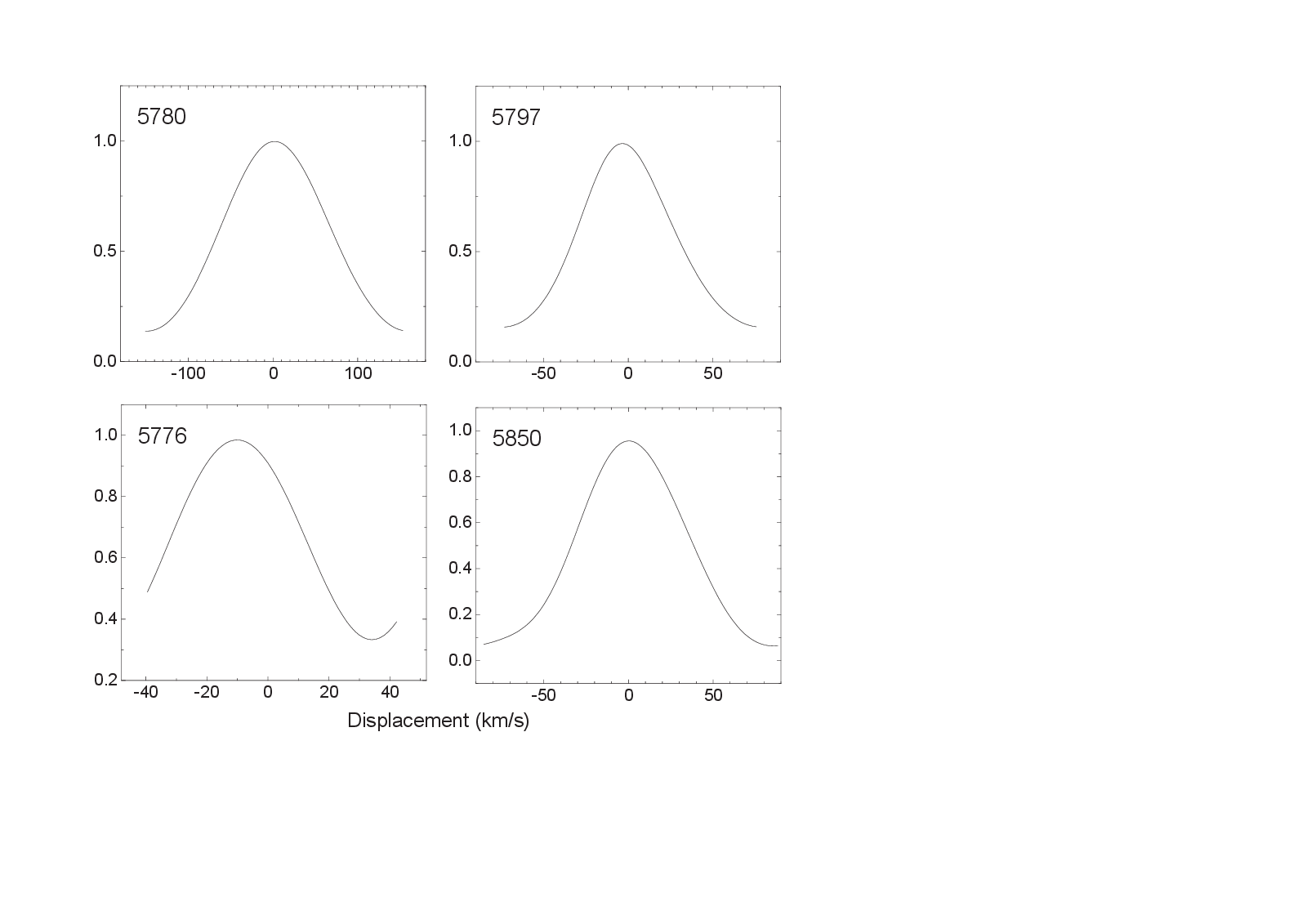}
  \caption{
  Cross-correlation functions of HD34078 diffuse bands shown in Fig. 6.
  }
}
\label{ccf}
\end{figure}

\begin{figure}
\center{\includegraphics[angle=0,width=14cm]{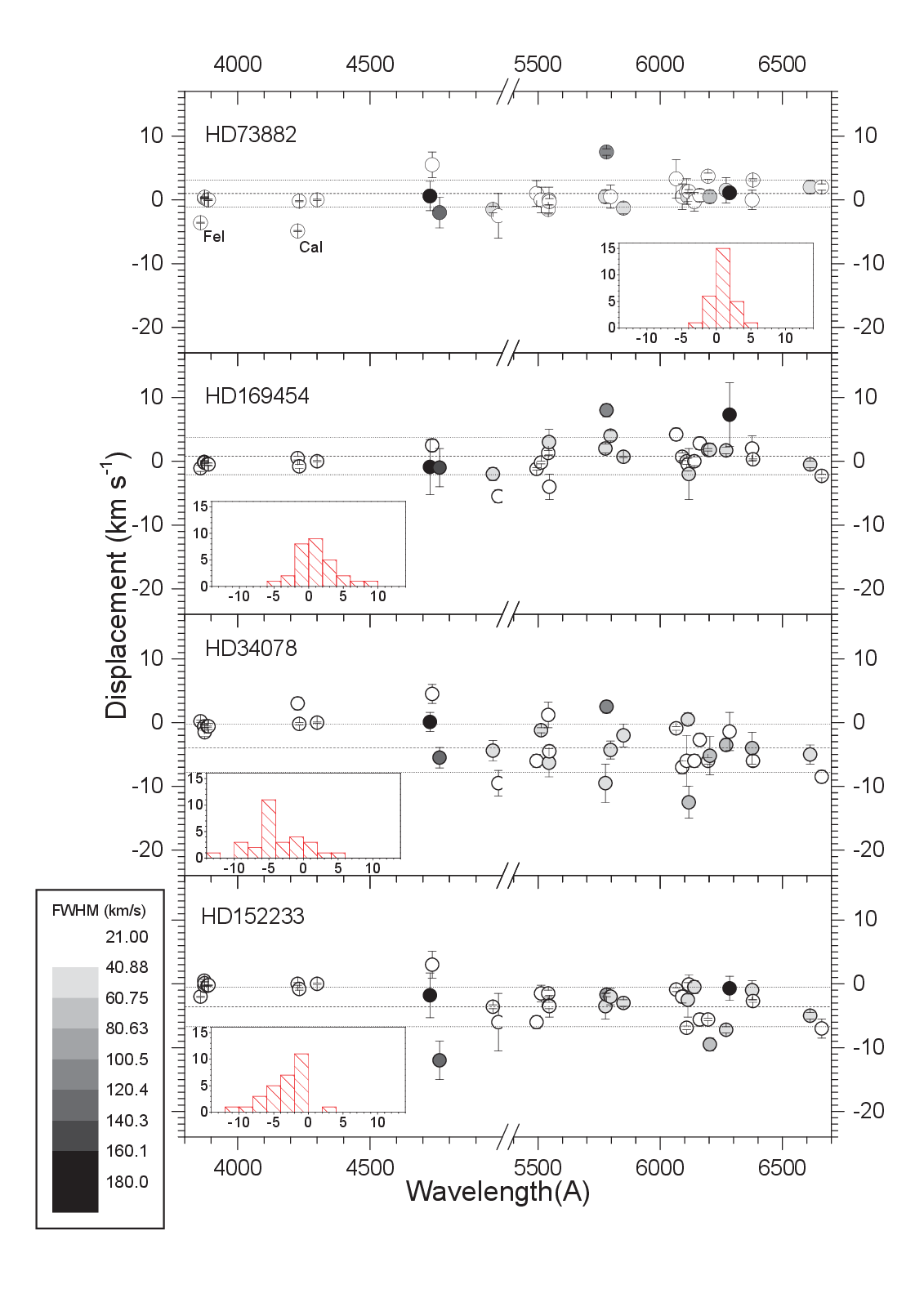}
  \caption{
  Positional displacement of interstellar features relative to the CH 4300~\AA\ band (see Tab. 3)
  and histograms of the distribution of displacements  (for DIBs only) for each studied object. Points with the wavelength lower than 4500 \AA\
  are well known molecules and atoms. Note the displaced Fe{\sc i} and Ca{\sc i} revealing tiny CaFe clouds (Bondar et al., 2007).
  Dash and dot lines represent the mean displacement and limits of the standard deviation for diffuse bands only.
  }
}
\label{scatter}
\end{figure}

\begin{figure}
\center{\includegraphics[angle=270,width=15cm]{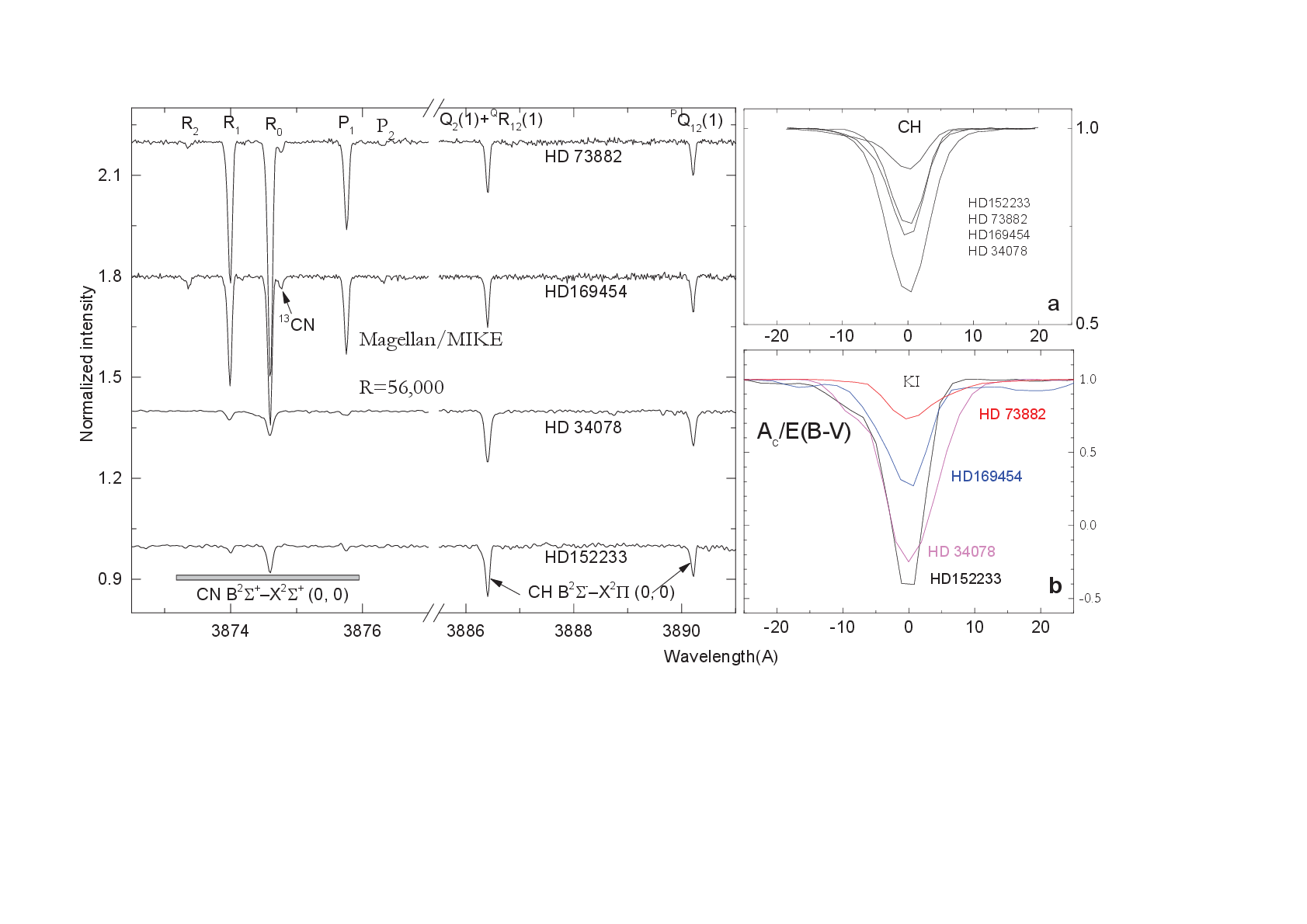}
 \caption{
 {\bf Left.} CN and CH B-X (0,0) violet interstellar bands. The features have been shifted to the rest wavelength velocity frame together with the whole spectra.
 The spectra are normalized to the identical depth of the CH 3886~\AA\ feature.
 {\bf Right. a } CH 4300 \AA\ line profiles in spectra of observed targets. Note the lack of Doppler split.
 {\bf Right. b } The potassium $\lambda$7699~\AA\ line, normalized to E(B-V) in the spectra of our targets. Apparently the ionization
 level of potassium is much lower in HD34078 and in HD152233 than in the two remaining targets. Note the lack of Doppler split.
 }
}
\label{compCHCN}
\end{figure}

\begin{figure}
\center{\includegraphics[angle=270,width=10cm]{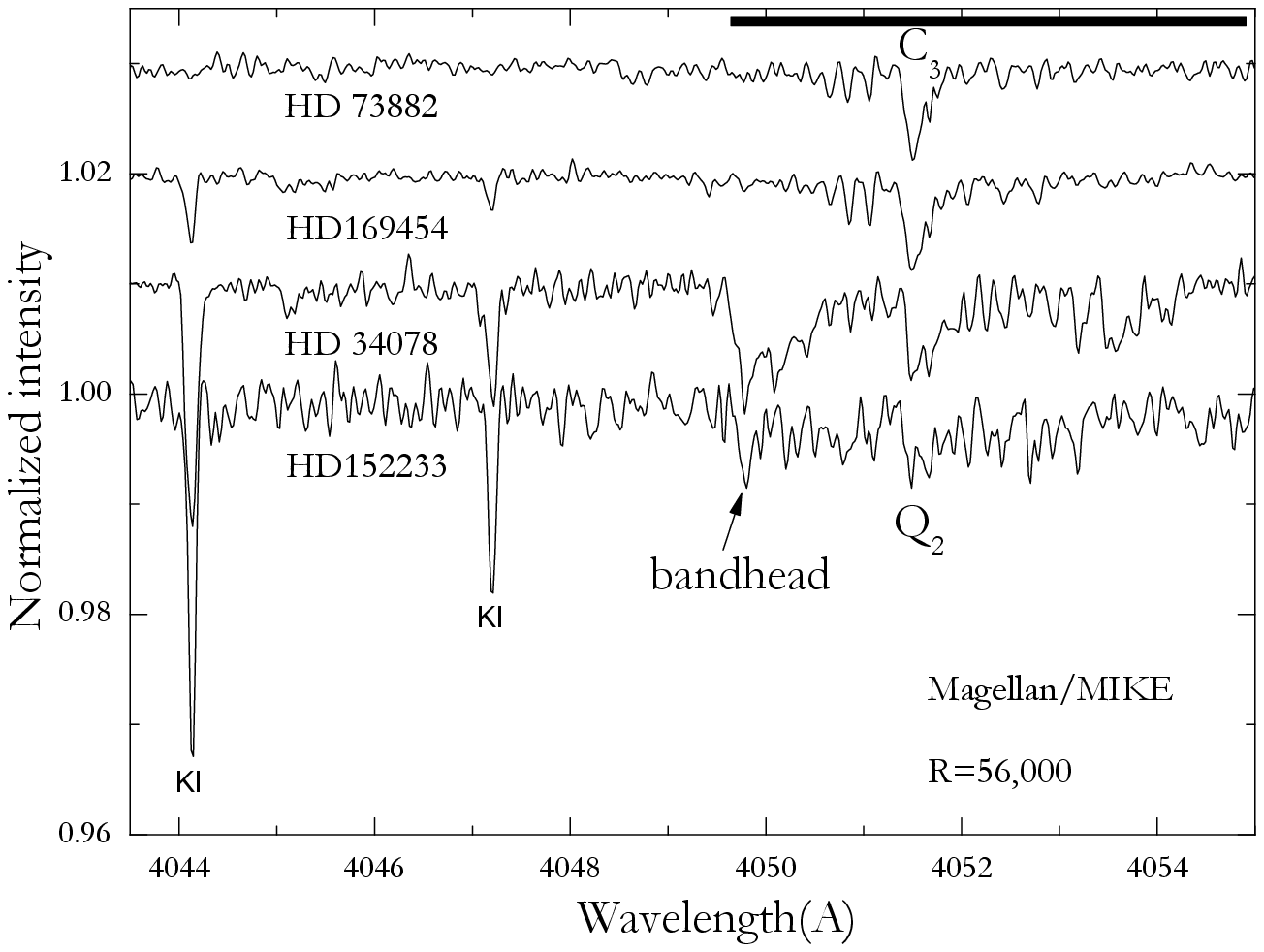}
\caption{The C$_3$ \~{A} $^{a}\Pi_{u}$ - \~{X}
$^{a}\Sigma_{g}^{+}$ 000-000 band  observed in all considered
targets.  Note the very strong bandheads in HD34078 and
HD152233 which demonstrate very high rotational temperatures. }
}
\label{compC3}
\end{figure}

\begin{figure}
 \center{\includegraphics[angle=270,width=90mm]{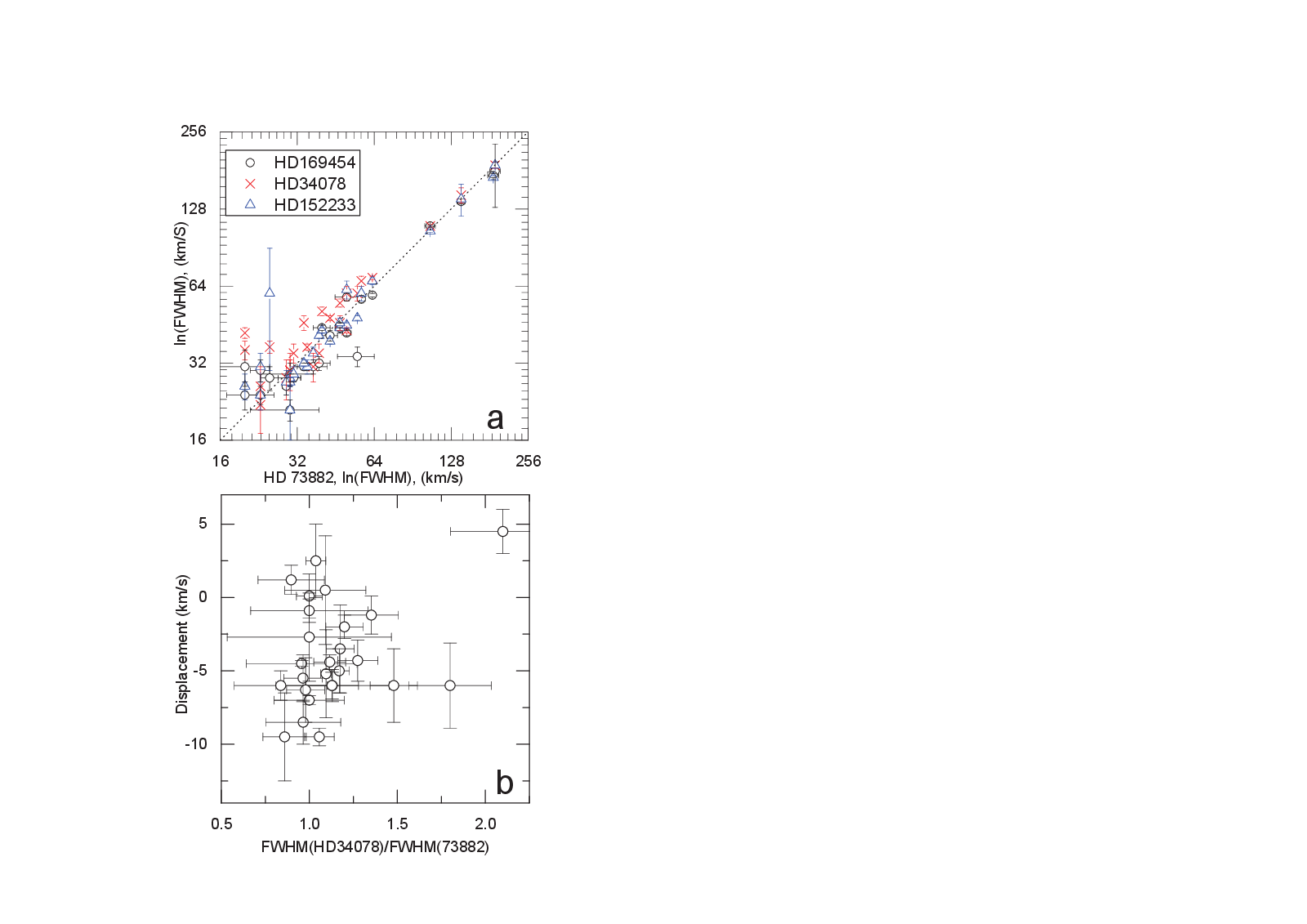}
 \caption{{\bf a.} Comparison of the width of diffuse interstellar bands measured in program stars.
 {\bf b.} The ratio of FWHM(HD34078)/FWHM(HD73882) versus the shift. Note the lack of correlation.}}
 \label{fwhm}
\end{figure}

\begin{figure}
\center{\includegraphics[angle=270,width=16cm]{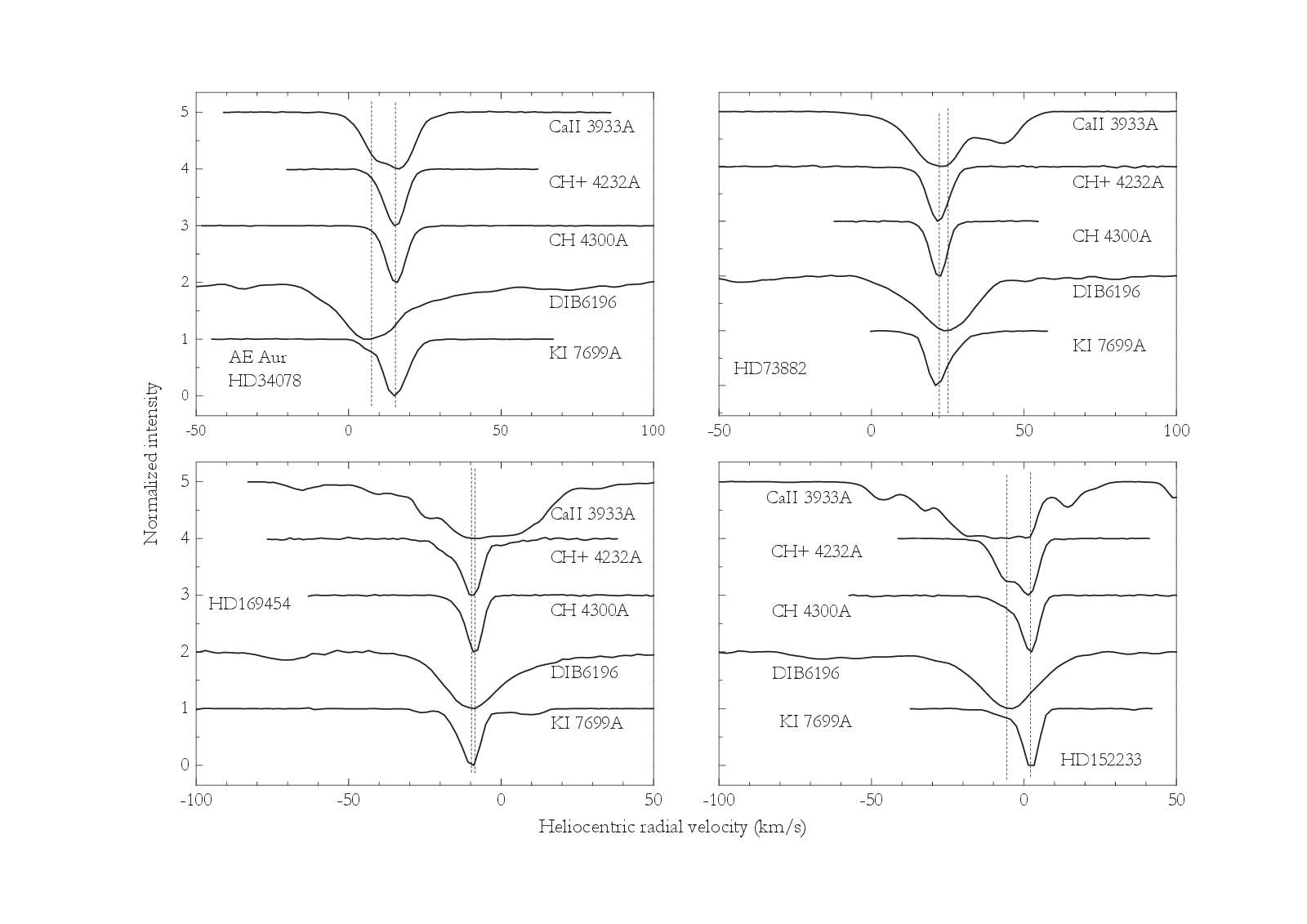}
  \caption{The identified interstellar features and the 6196 DIB plotted in the
  radial velocity scale in spectra of our targets.
  Vertical dotted lines mark position of CH 4300 and DIB6196.
  }
}
\label{panels4}
\end{figure}

\end{document}